\documentclass[11pt,tightenlines,eqsecnum,floats,aps,amsmath,amssymb,nofootinbib,prd,floatfix]{revtex4}

\usepackage{graphicx, wrapfig}
\usepackage{amssymb}
\usepackage{slashed}
\usepackage{bm}
\usepackage{color}
\usepackage{mathrsfs}
\usepackage{subfigure}
\usepackage{comment}
\usepackage{bold-extra}
\usepackage{hyperref}

\def\f{\frac}

\def\d{\textrm{d}}

\def\t{\tilde}

\def\pp{p_{\phi}}
\def\dpp{\delta p_{\phi}}
\def\pz{\mathring{\pi}}
\def\dph{\delta\phi}

\def\t{\tilde}
\def\h{\hat}

\newcommand{\be}{\nopagebreak[3]\begin{equation}}
\newcommand{\ee}{\end{equation}}
\def\hz{\mathring{h}}

\def\l{\left}
\def\r{\right}

\def\v{\vec}
\def\A1IJ{{\cal A}_1^{ij}}
\def\A2IJ{{\cal A}_2^{ij}}
\def\A3IJ{{\cal A}_3^{ij}}
\def\A4IJ{{\cal A}_4^{ij}}
\def\A5IJ{{\cal A}_5^{ij}}
\def\A6IJ{{\cal A}_6^{ij}}
\def\A1ij{{\cal A}^1_{ij}}
\def\A2ij{{\cal A}^2_{ij}}
\def\A3ij{{\cal A}^3_{ij}}
\def\A4ij{{\cal A}^4_{ij}}
\def\A5ij{{\cal A}^5_{ij}}
\def\A6ij{{\cal A}^6_{ij}}

\usepackage{enumerate}

\newcommand{\bfig}{\nopagebreak[3]\begin{figure}}
\newcommand{\efig}{\end{figure}}
\newcommand{\bea}{\nopagebreak[3]\begin{eqnarray}}
\newcommand{\ea}{\end{eqnarray}}

\newcommand{\bmult}{\nopagebreak[3]\begin{multline}}
\newcommand{\emult}{\end{multline}}

\usepackage{colordvi}
\usepackage{color}

\definecolor{grenn}{RGB}{0,103,0}

\newcounter{inout}
\newcommand{\inoutput}{%
	\stepcounter{inout}%
	\theinout}

\newcommand{\mathIn}[1]{
	\begin{quotation}\hspace*{-2.5cm} {\em In[\inoutput] :=\ } {\tt #1} \end{quotation}}

\newcommand{\mathOut}[1]{
	\begin{quotation} \hspace*{-2.5cm} {\em Out[\theinout] :=\ } #1 \end{quotation}}

\begin{document}

\title{xAct Implementation of the Theory of Cosmological Perturbation in Bianchi I Spacetimes\footnote{Contribution to the  Special Issue \href{https://www.mdpi.com/journal/mathematics/special_issues/computational_cosmology}{\it Mathematical and Computational Cosmology}, published by {\it Mathematics}, and edited by Prof. J. Velhinho.}}

\author{Ivan Agullo${}^{(1)}$}
\email{agullo@lsu.edu}
\author{Javier Olmedo${}^{(1)}$}
\email{javolmedo@ugr.es}
\author{V.~Sreenath${}^{(2)}$}
\email{sreenath@nitk.edu.in}
\affiliation{
${}^{(1)}$Department of Physics and Astronomy, Louisiana State University, Baton Rouge, LA 70803, U.S.A.
}
\affiliation{${}^{(2)}$Department of Physics, National Institute of Technology Karnataka, Surathkal, Mangalore 575025, India.}

\begin{abstract}
This paper presents a computational algorithm to derive the theory of linear gauge invariant perturbations on anisotropic cosmological spacetimes of the Bianchi I type. Our code is based on the tensor algebra packages {\tt xTensor} and {\tt xPert}, within the computational infrastructure of {\tt xAct} written in {\tt Mathematica}. The algorithm is based on a Hamiltonian, or phase space formulation, and it provides an efficient and transparent way of isolating the gauge invariant degrees of freedom in the perturbation fields and to obtain the Hamiltonian generating their dynamics. The restriction to Friedmann--Lema\^itre--Robertson--Walker spacetimes is straightforward.

\end{abstract}

\maketitle

\section{Introduction}
\label{sec:introduction}

Our understanding of the physics of the early universe is connected to cosmological perturbation theory. This framework describes the evolution of cosmological spacetimes, together with matter and gravitational perturbations propagating thereon. The description of perturbations  is technically complicated, mainly due to the fact that general relativity is a gauge theory. Due care is needed to separate gauge artifacts from physical effects. A natural and conceptually clean strategy is to work with gauge invariant fields, which are combinations of matter and gravitational perturbations that are left invariant under diffeomorphisms, or coordinate transformations, at the desired order in the perturbative expansion. This strategy was first implemented by Bardeen \cite{brdn} by expanding Einstein's equations to first order in perturbations, and it has been extensively used since then. An alternative and more geometric treatment can be obtained by working in the Hamiltonian, or phase space formulation of general relativity \cite{Arnowitt:1962hi}. Here, the phase space is equipped with four constraints---the so-called scalar and vector constraints---which are the generators of gauge transformations. Working at leading order in perturbations, one defines the gauge invariant fields as those combinations that remain unchanged under the transformations generated by the linearized constraints or, in simpler words, that Poisson-commute with them. Hence, at the practical level,  the task of finding gauge invariant perturbations reduces to finding combinations of linear perturbations whose Poisson brackets with the linearized constraints vanish. Furthermore, this procedure is equivalent to finding an appropriate canonical transformation, in such a way that the search for gauge invariant fields reduces to solving a Hamilton-Jacobi-like equation for the generating function of the transformation---with the added advantage that this equation becomes 
a set of simple algebraic equations for the unknown coefficients. This strategy was first implemented by Langlois \cite{lang} for Friedmann--Lema\^itre--Robertson--Walker (FLRW) cosmologies (see Refs. \cite{aan1,aan2,hyb1,hyb2} for a recent discussion), and it has been extended to anisotropic Bianchi I spacetimes in \cite{aos} (see also \cite{Pereira:2007yy,Pitrou:2008gk} for a previous analysis starting from Einstein equations).  

Although more systematic and geometric, the phase space derivation of linear cosmological perturbations is still extremely tedious and lengthy, in particular in the presence of anisotropies. In fact, the complexity of the calculations is one of the main barriers that researchers find when entering this field. The goal of this paper is to alleviate this issue by  introducing a computational algorithm to find gauge invariant linear perturbations in Bianchi I spacetimes, and the equations of motion they satisfy in the symbolic language of {\tt Mathematica}. Our algorithm is based on the 
package {\tt xPert} written for {\tt Mathematica}, which is embedded on the tensor algebra packages {\tt xTensor} of the {\tt xAct} distribution \cite{xAct,Brizuela:2006ne,Brizuela:2008ra} (they are available under the General Public License). A previous, and more general implementation of perturbation theory using similar tools was published in \cite{Pitrou_2013}; it is more general because it applies to all orders in perturbation theory and to any spatially homogenous spacetimes in any gauge---it does not deal, however, with gauge invariant perturbations.  (See also \cite{Brizuela:2009qd,Brizuela:2010qu} for the application to perturbations on spherically symmetric spacetimes.) 
The merit of our algorithm is that we combine a series of analytical and computational techniques to make the problem of finding gauge invariant variables at leading order in perturbations tractable. We find that the phase space formulation  helps enormously  in these respects and, when combined with carefully thought numerical algorithms, 
 it gives rise to a computational infrastructure of great utility to researchers in this field. This manuscript is accompanied by a {\tt Mathematica} notebook, publicly available at \cite{ntbk}, that contains a step by step implementation of the algorithms presented here. 
Furthermore, we have complemented the analysis  with another code, based on the C programming language and publicly available in  \cite{num-lib}, to solve the equations of motions of gauge invariant perturbations and to compute observable quantities in the cosmic microwave background (CMB) starting from suitable initial data, although the details of this code are not spelled out in this paper. 
  

This article is organized in such a way that the  procedure has been separated in small steps that are, one by one,  introduced theoretically and  implemented computationally. Namely,  section \ref{sec:background} introduces  Bianchi I spacetimes and linear perturbations thereon in the Arnowit-Deser-Misner (ADM) formalism \cite{Arnowitt:1962hi}. Section \ref{sec:SVT} introduces the generalization of the scalar-vector-tensor decomposition commonly used in FLRW backgrounds. Section \ref{sec:gaugeinvarinat} introduces gauge invariant fields, and in section \ref{sec:2nd-H} we study their dynamics. Section \ref{sec:concl} contains a short summary and some concluding remarks.


\section{Perturbed Bianchi I spacetimes in the 
ADM formalism}\label{sec:background}

\subsection{Summary of the theory}

We start with Einstein's gravity minimally coupled to a scalar field $\Phi$ with potential energy density $V(\Phi)$ and no anisotropic stresses. We assume  the spacetime manifold to be $M=\mathbb{R} \times M_3$, where $M_3$ has $\mathbb{R}^3$  topology. In the following, we will restrict ourselves to a finite volume $\mathcal{V}_0$ relative to an auxiliary flat Euclidean metric $\delta_{ij}$ defined in $M_3$.\footnote{This is equivalent to put the universe in a box of arbitrarily large but finite volume $\mathcal{V}_0$, with periodic boundary conditions. We do this for convenience in the expressions below. The volume $\mathcal{V}_0$ will not impact  predictions, and it can be taken to infinity at the end of the calculations.} 
We will adopt a Hamiltonian formulation following ADM \cite{Arnowitt:1962hi}. In this formalism, elements of the phase space are made of four real fields $(\Phi(\vec{x}),P_{\Phi}(\vec{x}),h_{ij}(\vec{x}),\pi^{ij}(\vec{x}))$ defined in $M_3$, where Latin indices $i,j$ run from 1 to 3. Here, $h_{ij}(\vec{x})$ is a Riemannian metric that describes the intrinsic spatial geometry of $M_3$, and $\pi^{ij}(\vec{x})$ its conjugate momentum. The non-vanishing Poisson brackets between these fields are
\be \label{PB} \{ \Phi(\vec{x}),P_{\Phi}(\vec{x}')\}=\delta^{(3)}(\vec{x}-\vec{x}')\, ,\hspace{1cm}
\{ h_{ij}(\vec{x}),\pi^{k l}(\vec{x}')\}=\delta_{(i}^k\delta_{j)}^{l}\delta^{(3)}(\vec{x}-\vec{x}')\, .\ee
where $\delta_{(i}^k\delta_{j)}^{l}\equiv \frac{1}{2} (\delta_{i}^k\delta_{j}^{l}+\delta_{j}^k\delta_{i}^{l})$. Dynamics is generated by the Hamiltonian 
\be \label{ham} \mathcal H=\int \d^3x \, \Big[N(\vec{x}) \,  \mathbb S (\vec{x})+N^i(\vec{x}) \, \mathbb V_i (\vec{x})\Big]\, ,\ee
which is a combination of first class constraints, and $N(\vec{x})$ and $N^i(\vec{x})$ (the lapse and shift, respectively) play the role of Lagrange multipliers. Concretely, $\mathbb S (\vec{x})$ is the scalar constraint and $ \mathbb V_i (\vec{x})$ are the vector or diffeomorphism constraints. In terms of the canonical variables, they have the form
\bea \label{scons} \mathbb S (\vec{x})&=&\f{2\kappa}{\sqrt{h}}\l( \pi^{ij}\pi_{ij} -\f{1}{2}\pi^2 \r) - \f{\sqrt{h}}{2\kappa} ~^{(3)}R + \f{1}{2\sqrt{h}} P_{\Phi}^2 + \sqrt{h} \, V(\Phi) + \f{\sqrt{h}}{2} D_i \Phi D^i \Phi \approx 0 \, ,  \\ \label{vcons}
\mathbb V_i(\vec{x})&=&  -2 \sqrt{h}\,  h_{ij} \, D_k(h^{-1/2} \pi^{kj}) +P_{\Phi} \, D_i \Phi \approx 0\, , \ea
where $\kappa=8\pi G$, and $D_i$ is the covariant derivative compatible with the spatial metric $h_{ij}$, $h$ its determinant, and $^{(3)}R$ its Ricci scalar curvature. Given $N(\vec{x})$, $N^i(\vec{x})$, and a solution to Hamilton's equations, $h_{ij}(\vec{x},t)$, the spacetime metric takes the form 
\begin{equation}
 \d s^2 = -(N^2 - N_iN^i)\,  \d t^2 + 2 N_i \, \d x^i \d t + h_{ij}\,  \d x^i \d x^j,
\end{equation}
where $t$ is a time variable that labels each space-like hyper-surface $M_3(t)$, and $x^i$ are spatial coordinates on them. 

Let us now focus on the sector of the phase space of general relativity that is made of Bianchi I geometries together with small inhomogeneous perturbations. This  is commonly done by considering  curves $\gamma[\epsilon]$ in the ADM  phase space that pass through the Bianchi I subspace at $\epsilon=0$. Expanding the phase space variables around $\epsilon=0$, we have: 
\bea \label{expan} h_{ij}(\vec{x},\epsilon) &=& \hz_{ij} + \epsilon\,
\delta h^{(1)}_{ij}(\vec{x}) + \ldots + \f{\epsilon^n}{n!}\, 
\delta h_{ij}^{(n)}(\vec{x}) + \ldots\, ,
\nonumber\\
\pi^{ij}(\vec{x},\epsilon)&=&\pz^{ij}+\epsilon\,
\delta \pi^{ij\, (1)}(\vec{x}) + \ldots + \f{\epsilon^n}{n!}\, 
\delta\pi^{ij\, (n)}(\vec{x}) + \ldots\, , \nonumber \\
\Phi(\vec{x},\epsilon) &=& \phi + \epsilon\, \dph^{(1)}(\vec{x}) + \ldots  + \f{\epsilon^n}{n!}\,\dph^{(n)}(\vec{x}) + \ldots\, , \nonumber \\
P_{\Phi}(\vec{x},\epsilon)&=&\pp+\epsilon\ \dpp^{(1)}(\vec{x})+ \ldots+\f{\epsilon^n}{n!}\,\dpp^{(n)}(\vec{x}) + \ldots\, ,
 \ea
where $\phi$, $\pp$, $\hz_{ij}$, $\pz^{ij}$ describe a Bianchi I background geometry,  and $\dph^{(n)}(\vec{x})$, $\dpp^{(n)}(\vec{x})$, $\delta h^{(n)}_{ij}(\vec{x})$, $\delta \pi^{ij\, (n)}(\vec{x})$ describe  $n$th-order perturbations thereon.  The homogeneous variables satisfy the  Poisson brackets
\be \label{bpb}  \{\phi, \pp\}=\frac{1}{\mathcal{V}_0}\, ,  \hspace{0.5cm} \{\hz_{ij}, \pz^{kl} \}=\frac{1}{\mathcal{V}_0} \, \delta_{(i}^k\delta_{j)}^{l}\, .\ee 
As it is common in the literature, we restrict ourselves to diagonal Bianchi I metrics, such that the phase space variables $\hz_{ij}$ and $\pz^{ij}$ take a diagonal form in an appropriate system of coordinates $x^i$
\be \label{back} \hz_{ij}={\rm diag}(a_1^2,\,a_2^2,\,a_3^2) \, ,  \hspace{1cm} \pz^{ij}={\rm diag}\left(\f{\pi_{a_1}}{2\, a_1},\, \f{\pi_{a_2}}{2\, a_2},\, \f{\pi_{a_3}}{2\, a_3}\right)  \, ,\ee
where $a_i$ define the three directional scale factors; it follows from (\ref{bpb})  that $a_i$ and $\pi_{a_j}$ are canonically conjugate, $\{a_i,\pi_{a_j}\}=\frac{1}{\mathcal{V}_0} \delta_{ij}$ (note that the subscripts  $i,j$ in $a_i$ and $\pi_{a_j}$ are just labels, and not tensorial indices). From now on we will raise and lower all spatial indices $i,j,k,...$ with  $\hz_{ij}$ and its inverse. 

The next step is to expand the constraints (\ref{scons}) and (\ref{vcons}) in perturbations 
\bea \mathbb S(\vec x)&=&\mathbb S^{(0)}+\mathbb S^{(1)}(\v x)+\mathbb S^{(2)}(\v x)+\mathbb S^{(3)}(\v x)+\cdots\,  ,\nonumber \\
 \mathbb V_i(\v x)&=&\mathbb V_i^{(0)}+\mathbb V_i^{(1)}(\v x)+\mathbb V_i^{(2)}(\v x)+\mathbb V_i^{(3)}(\v x)+\cdots\, ,\ea
where the superscripts in parenthesis denote the order in our perturbative expansion. As mentioned before, we want to focus on linear perturbations. This will require, on the one hand, to keep only first order perturbations $\delta h^{(1)}_{ij}(\vec x)$, $\delta \pi^{ij\, (1)}(\vec{x})$, $\dph^{(1)}(\vec{x})$, and $ \dpp^{(1)}(\vec{x})$---since these will be the only perturbations in the rest of this paper, from now on we will  remove the label $(1)$---and, on the other hand, to truncate all constraints at quadratic order in these fields. 

In addition to the perturbative expansion of the constraints, we also expand  the lapse function as $N+\delta N(\vec x)$, and the shift as $N^i+\delta N^i(\vec x)$, where, in the following, $N$ is a homogeneous function, and we take $N^i=0$, so the background line element takes the familiar form
\begin{equation}
 \d s^2 = -N^2\d t^2 + \hz_{ij}\,  \d x^i \d x^j.
\end{equation}

In the rest of this section we  discuss the background sector, and leave the study of the inhomogeneous degrees of freedom for the next section. Because of homogeneity,  $\mathbb V_i^{(0)}$ identically vanish (note that they are proportional to derivatives in space-like directions). Hence, the homogeneous degrees of freedom are subject to only one constraint, $\mathbb S^{(0)}$, which takes the form 
\bea  \label{Fcons} \mathbb S^{(0)}&=& \f{1}{2\sqrt{\hz}} \biggl[ 
 \kappa \l( \f{a_1^2 \pi_{a_1}^2}{2} + \f{a_2^2 \pi_{a_2}^2}{2}  + \f{a_3^2 \pi_{a_3}^2}{2} 
 - a_1 \pi_{a_1} a_2 \pi_{a_2} - a_2 \pi_{a_2} a_3 \pi_{a_3} - a_3 \pi_{a_3} a_1 \pi_{a_1} \r)\nonumber\\
 && + p_{\phi}^2 + 2 \hz V(\bar{\phi}) \biggr] \approx 0,
\ea%
with  $\hz=(a_1a_2a_3)^2=a^6$ the determinant of $\hz_{ij}$, and we have defined $a\equiv (a_1a_2a_3)^{1/3}$ as the mean scale factor. Then, the Hamiltonian (\ref{ham})  reduces to
\be \label{backH0} \mathcal H_{_{\rm BI}}=\int_{M_3} \d^3x \, N \, \mathbb S ^{(0)}=\mathcal{V}_0\, N \, \mathbb S ^{(0)}\, .
\ee
If we choose $N=1$, $\mathcal H_{_{\rm BI}}$ generates evolution in standard cosmic time $t$. Hamilton's equations of motion then read 
\bea \label{eoma} \dot a_i&=&\{a_i, \mathcal H_{_{\rm BI}}\} ,\hspace{0.5cm} \dot \pi_{a_i}=\{\pi_{a_i}, \mathcal H_{_{\rm BI}}\} \, , \\ \nonumber 
\label{eomphi} \dot \phi&=&\{\phi, \mathcal H_{_{\rm BI}}\} \, , \hspace{1cm} \dot{\pp}=\{\pp, \mathcal H_{_{\rm BI}}\} \, .\ea
These equations  fully determine the dynamical evolution of the Bianchi I geometry, once suitable initial data satisfying the scalar constraint is provided. For convenience in the interpretation of the solutions, it is useful to introduce (see e.g.\ Ref. \cite{aos} for details), on the one hand, the average Hubble rate $H=\frac{\dot a}{a}$. Its relation  to the directional Hubble rates $H_i\equiv \frac{\dot a_i}{a_i}$ is  $H = \frac{1}{3}\left(H_1+H_2+H_3\right)$. On the other hand, the anisotropic shear $\sigma_{ij}$ is defined from $\pz_{ij}$ by
\be\label{pisigma}  \pz_{ij}= \frac{h^{1/6}}{6}\, \pi_a\,  \hz_{ij}+\frac{h^{1/2}}{2\kappa}\, \sigma_{ij} \, , \ee  
where $\pi_a$ is the conjugate momenta of $a$ (its can be written in terms of  $\dot a$ as $\pi_a=-\frac{6}{\kappa}\, a \, \dot a$). Equation (\ref{pisigma}) is equivalent to saying that the components of $\sigma_{ij}$
\be \sigma_{ij}={\rm diag}(a_1^2\, \sigma_1,a_2^2\, \sigma_2,a_3^2\, \sigma_3)\, ,\ee
are related to the canonical variables $\pi_{a_i}$ by 
\be \pi_{a_i}=\frac{1}{\kappa}\frac{a^3}{a_i}\, (\sigma_i-2\, H)\, . \ee
Using the equations of motion (\ref{eoma}), one can check that $\sigma_i=H_i-H$, and from this it is obvious that  $\sigma_i$ are not all independent, but   satisfy  $\sigma_1+\sigma_2+\sigma_3=0$ --- or in other words, $ \sigma_{ij}$ is traceless.
 It is also convenient to define the shear squared 
\be\sigma^2=\sigma_{ij}\sigma^{ij}=\sigma_1^2+\sigma_2^2+\sigma_3^2=(H_1-H)^2+(H_2-H)^2+(H_3-H)^2\, .
\ee
With these  definitions, the equations of motion (\ref{eoma}) can be written in the  more familiar form
\be \label{aver} \frac{\ddot a}{a}=-\frac{\kappa}{6}\, [\rho+3\, P]-\frac{\sigma^2}{3} \, , \hspace{1cm} \ddot \phi+3\f{\dot a}{a}\, \dot\phi+\frac{\d V(\phi)}{\d\phi}=0 \, , \ee
and the scalar constraint (\ref{Fcons}) as
\be  \label{cons} H^2=\frac{\kappa}{3}\, \rho +\frac{\sigma^2}{6}\, . \ee
Here, $\rho\equiv \f{1}{2}\dot \phi^2+V(\phi)$ and $P\equiv \f{1}{2}\dot \phi^2-V(\phi)$ are the energy and pressure densities of $\phi$, respectively.  The expressions above are equivalent to the diagonal components of Einstein's equations. For $\sigma^2=0$ they reduce to the Friedmann-Lema\^itre theory of  isotropic cosmologies. On the other hand, the equations of motion for the shear are
\be \label{eqanis} \dot \sigma^i_{\  j}=-3\, H\, \sigma^i_{\ j} \, , \ee
whose solutions are simply 
$\sigma_i=\Sigma_i/a^3$, where $\Sigma_i$ are constants constrained by $\Sigma_1+\Sigma_2+\Sigma_3=0$. This implies $\sigma^2=\frac{\Sigma^2}{a^6}$, with $\Sigma^2\equiv \Sigma_1^2+\Sigma_2^2+\Sigma_3^2$.

\subsection{Implementation in {\tt Mathematica}}

We will begin here the description of the  main steps carried out in the {\tt Mathematica} notebook \cite{ntbk}.

\subsubsection{Preliminaries}

The perturbative expansions are carried out employing the package {\tt xPert}~\cite{Brizuela:2008ra}.  Hence, the first step is to load this package with the following command
\mathIn{<<xAct`xPert`}
We define the three-dimensional manifold {\tt M3} with abstract indices $\{ {\tt i, j, k, l, m, n} \}$: 
\mathIn{DefManifold[M3,3,$\{{\tt i, j, k, l, m, n}\}$];}
The spatial slices will be parameterized by a time variable $\tt t$. We define it with the command
\mathIn{DefParameter[t,PrintAs->"$\tt t$"];}
We also define the gravitational coupling constant:
\mathIn{DefConstantSymbol[$\kappa$];}
The (Riemannian) spatial metric {\tt h}  and its covariant derivative {\tt CD} are introduced by means of
\mathIn{DefMetric[1,h[-$\tt i$,-$\tt j$],CD,\{";","$\tt D$"\}, Otherdependencies->$\{\tt t\}$,WeightedWithBasis->AIndex];}
Note that we have allowed the spatial metric $\tt h$ to depend on the time parameter $\tt t$. We now introduce perturbations of the metric:
\mathIn{DefMetricPerturbation[h,$\delta$h,$\epsilon$];}
We define the scalar field:
\mathIn{DefTensor[$\phi$[],\{M3,t\}];}
and its potential:
\mathIn{DefScalarFunction[V];}
We incorporate perturbations of the scalar field with
\mathIn{DefTensorPerturbation[$\delta\phi$[LI[1]],$\phi$[],\{M3,t\}];}
We define next canonical momenta. The momentum conjugate to the spatial metric is introduced as:
\mathIn{DefTensor[P[i,j],\{M3,t\}];}
and its perturbations defined by
\mathIn{DefTensorPerturbation[$\delta$P[LI[1],i,j],P[i,j],\{M3,t\}];}
Finally, we define the momentum conjugate to the scalar field
\mathIn{DefTensor[P$\phi$[],\{M3,t\}];}
and its corresponding perturbation:
\mathIn{DefTensorPerturbation[$\delta$P$\phi$[LI[1]],P$\phi$[],\{M3,t\}];}

\subsubsection{Scalar and vector constraints}

We start introducing the diffeomorphism constraints defined in (\ref{vcons}):
\mathIn{diffeo=-2PD[-k]@(h[-i,-j]P[j,k])+P[k,j]PD[-i]@h[-k,-j]+P$\phi$[]PD[-i]@$\phi$[];}
As we will see below, only the linear term in the perturbative expansion of these constraints, called $\mathbb V_i^{(1)}(\v x)$ above, will be relevant in the description of linearized perturbations (recall also that $\mathbb V_i^{(0)}$ are identically zero). They are defined in the notebook by:
\mathIn{(Perturbed[diffeo,1]/$\epsilon$);\\
diffeo1\,=\,\%\,/.MakeRule[{PD[-i]@$\phi$[],0}]/.MakeRule[{PD[-i]@h[-j,-k],0}]\\
/.MakeRule[{PD[-i]@P[j,k],0}];}
In this  expression, we have imposed that the partial spatial derivatives of the background degrees of freedom vanish because of homogeneity. 

Let us focus now on the scalar constraint. We are going to compute each of its terms, written in (\ref{scons}), separately. First of all,  the three-dimensional Ricci curvature is
\mathIn{ricci=(Deth[])]$^{\wedge}$(1/2)h[j,k]RiemannCD[-j,-i,-k,i]//RiemannToChristoffel\\//ChristoffelToMetric//Simplification//NoScalar;}
We now expand this term in perturbations by using
\mathIn{Perturbed[ricci,2]/.MakeRule[\{PD[-i]@h[-j,-k],0\}]//ExpandPerturbation;
  \\r2\,=\,\%/.MakeRule[{h[LI[2],-i,-j],0}];}
where we have  imposed again  homogeneity of the background metric, $\partial_ih_{jk}=0$, and we have put to zero the second order perturbations, $\delta h^{(2)}_{ij}=0$. 

On the other hand, the first term in (\ref{scons}) (the ``kinetic''  term of the gravitational sector)  is 
\mathIn{(Deth[])$^{\wedge}$(-1/2)P[i,j]P[k,l](h[-i,-k]h[-j,-l]-h[-i,-j]h[-k,-l]/2);\\
ExpandPerturbation[Perturbed[\%,2]];\\
pipi\,=\,\%\,/.MakeRule[\{$\delta$P[LI[2],-i,-j],0\}]/.MakeRule[\{$\delta$h[LI[2],-i,-j],0\}];}
where we have imposed $\delta h^{(2)}_{ij}=0$ and $\delta \pi^{(2)}_{ij}=0$ in the last line.

The terms in (\ref{scons}) that depend on the scalar field are
\mathIn{1/2Deth[]$^{\wedge}$(-1/2)P$\phi$[]$^{\wedge}$2+Deth[]$^{\wedge}$(1/2)(1/2PD[-i]@$\phi$[]PD[-j]@$\phi$[]h[i,j]\\
+V[$\phi$[]]);\\
ExpandPerturbation[Perturbed[\%,2]];\\
matter\,=\,\%\,/.MakeRule[\{$\delta$h[LI[2],-i,-j],0\}]/.MakeRule[\{$\delta\phi$[LI[2]],0\}]\\
/.MakeRule[\{$\delta$P$\phi$[LI[2]],0\}]/.MakeRule[\{PD[-i]@$\phi$[],0\}];}
Putting everything together, the scalar constraint, up to second order in perturbations, is
\mathIn{ S\,=\,Series[(2$\kappa$)pipi-1/(2$\kappa$)r2+matter,\{$\epsilon$,0,2\}];}
The scalar constraint contributes with 0th, 1st and 2nd order terms in the perturbative expansion. Let us identify each one. We first introduce the shear tensor as
\mathIn{DefTensor[$\sigma$[i,j],\{M3,t\},Symmetric[\{i,j\}]];}
and the shear  $\sigma\equiv\sqrt{\sigma^2}$
\mathIn{DefTensor[$\sigma$b[],\{M3,t\}];}
We impose that the shear is traceless and symmetric, and its relation with {\tt $\sigma$b[]$^{\wedge}$2}, with the following automatic rules
\mathIn{AutomaticRules[$\sigma$,MakeRule[\{$\sigma$[i,j]h[-i,-j],0\}]];\\
AutomaticRules[$\sigma$,MakeRule[\{$\sigma$[i,-i],0\}]];\\
AutomaticRules[$\sigma$,MakeRule[\{$\sigma$[i,j]$\sigma$[k,l]h[-i,-k]h[-j,-l],$\sigma$b[]$^{\wedge}$2\}]];}
In addition, the conjugate variable $\pi_a$ to the average scale factor $a$ is defined as
\mathIn{DefTensor[$\pi$a[],\{M3,t\}];}
Expression (\ref{pisigma}) above is implemented as
\mathIn{bgmomrule=MakeRule[\{P[i,j],$\pi$a[]/6 Deth[]$^{\wedge}$(1/6) h[i,j]+Deth[]$^{\wedge}$(1/2)/(2$\kappa$) $\sigma$[i,j]\}];}
With this, we express the first-order diffeomorphism constraints in terms of the shear as
\mathIn{diffeo1/.bgmomrule//org//ChristoffelToMetric//Simplification//NoScalar;\\
diffeoa=\%/.MakeRule[\{PD[-i]@h[-j,-k],0\}];}
Similarly, the 0th order scalar constraint is
\mathIn{S0\,=\,SeriesCoefficient[S,0];}
We write it in terms of $\sigma^{2}$ and $\pi_a$ by
\mathIn{S0a=S0/.bgmomrule//ToCanonical;}
In a similar way, we define the part of the scalar constraint that is linear in perturbations as
\mathIn{S1a\,=\,SeriesCoefficient[S,1];}
and in terms of shear:
\mathIn{S1b\,=\,S1a/.bgmomrule//ToCanonical;}
The  part of the scalar constraint that is quadratic in perturbations will be discussed in Sec. \ref{sec:2nd-H}.

\section{Scalar-Vector-Tensor decomposition} \label{sec:SVT}
\subsection{Summary of the theory}

Linear  perturbations satisfy, via  (\ref{PB}) and (\ref{bpb}), the canonical Poisson brackets
\be \label{pertPB} \{ \dph (\vec{x}),\dpp (\vec{x}')\}=\delta^{(3)}(\vec{x}-\vec{x}')-\frac{1}{\mathcal{V}_0}\, 
;\hspace{0.5cm} \{ \delta h_{ij}(\vec{x}),\delta \pi^{k l}(\vec{x}')\}=\delta_{(i}^k\delta_{j)}^{l}\, \Big(\delta^{(3)}(\vec{x}-\vec{x}')-\frac{1}{\mathcal{V}_0}\Big)\, .\ee
For convenience, we Fourier expand these perturbations and their conjugate  momenta\footnote{The Fourier expansion of fields is adapted to the fiducial cell of volume $\mathcal{V}_0$, so the wavenumbers $\vec k$ will take values on a discrete lattice $\vec k\in 2\pi/(\mathcal{V}_0)^{1/3}\, \mathbb{Z}^3$. In the limit $\mathcal{V}_0\to\infty$ one recovers $\vec k\in \mathbb{R}^3$.}
\be \delta \phi(\vec{x})=\sum_{\vec{k}\neq \v0} \delta \t{\phi}(\vec{k})\, e^{i\, \v{k}\cdot\v{x}}\, ; \hspace{1cm} \delta p_{\phi}(\vec{x})=\sum_{\vec{k}\neq \v 0} \delta \t{p}_{\phi}(\vec{k})\, e^{i\, \v{k}\cdot\v{x}}\, ,\ee

\be \label{Fexph} \delta h_{ij}(\vec{x})=\sum_{\vec{k}\neq \v 0} \delta \t{h}_{ij}(\vec{k})\, e^{i\, \v{k}\cdot\v{x}}\, ; \hspace{1cm} \delta \pi^{ij}(\vec{x})=\sum_{\vec{k}\neq \v 0} \delta \t{\pi}^{ij}(\vec{k})\, e^{i\, \v{k}\cdot\v{x}}\, ,\ee %
where $\v{k}\cdot\v{x}=k_i\,x^i$ and such that $k_i$ is time independent (the comoving wavevector). 

The Poisson brackets (\ref{pertPB}) become
 \be\label{cpb} \{ \delta  \t \phi(\vec{k}),\delta \t p_{\phi}(\vec{k'})\}=\mathcal{V}_0^{-1}\, \delta_{\vec{k},-\vec{k}'}\,; \quad \{ \delta  \t h_{ij}(\vec{k}),\delta \t \pi^{k l}(\vec{k'})\}=\mathcal{V}_0^{-1}\, \delta_{(i}^k\delta_{j)}^{l}\,  \delta_{\vec{k},-\vec{k}'}\, .\ee

We now perform a generalization of the  scalar-vector-tensor decomposition of $\delta \t{h}_{ij}(\vec{k})$ and $\delta \t{\pi}^{ij}(\vec{k})$ that is commonly used in FLRW spacetimes. Although this decomposition is adapted to the rotational invariance of FLRW geometries, it will also be useful in  Bianchi I, since it will allow us to work with variables that become the familiar scalar, vector, and tensor modes when the background geometry isotropizes (as it quickly happens if there is a phase of inflation). We define now a basis of $3\times 3$ symmetric matrices as
\begin{align}
 {A}^{{(1)}}_{ij}\, &=\, \f{\hz_{ij}}{\sqrt{3}}, \hspace{0.5in} & {A}^{(4)}_{ij}\, &=\,\f{1}{\sqrt{2}}\, \l(\, \hat{ k}_i\, \h e_{2 j}\, +\, \hat{ k}_j\, \h e_{2 i} \,\r),\nonumber\\
 {A}^{(2)}_{ij}\, &=\,\sqrt{\f{3}{2}}\,\l(\hat{ k}_i\,\hat{ k}_j - \f{\hz_{ij}}{3}\r), \hspace{.5in}  &{A}^{(5)}_{ij}\,& =\, \f{1}{\sqrt{2}}\, \l(\, \hat e_{1 i}\, \h e_{1 j}\, -\, \hat e_{2 i}\, \h e_{2 j} \,\r),  \nonumber\\
{A}^{(3)}_{ij}\, &=\, \f{1}{\sqrt{2}}\, \l(\, \hat{ k}_i\, \h e_{1 j}\, +\, \hat{ k}_j\, \h e_{1 i}\,\r),
 \hspace{.5in} &{A}^{(6)}_{ij}\, &=\,\f{1}{\sqrt{2}}\, \l(\, \hat e_{1 i}\, \h e_{2 j}\, +\, \hat e_{1 j}\, \h e_{2 i} \,\r). \label{matrixbases}
\end{align}
Here, $\hat{ k}$ is the unit vector (with respect to $\hz_{ij}$) in the direction of $\v{k}$, and $\h e_{1}$, $\h e_{2}$ are two  unit vectors orthogonal among themselves and  to $\hat{ k}$.\footnote{Note that the three unit vectors $\hat{k},\h e_{1}, \h e_{2}$ are time dependent. This is because, on the one hand, the norm of $k_i$ is time dependent and, on the other hand, the unit vectors $\h e_{1}, \h e_{2}$ need to rotate in time to remain orthogonal to $\hat{k}$, unless $\hat{k}$ points in one of the principal directions. For the details on how to compute the time dependence of the unit vectors $\hat{ k}$, $\h e_1$ and $\h e_2$, see Appendix A in Ref.\ \cite{aos}.} We now define $\gamma_n(\vec k)$ and $\pi_n(\vec k)$ as the components of $\delta \tilde h_{ij}(\vec{k})$ and $\delta \tilde \pi_{ij}(\vec{k})$, respectively, in this basis
 \be \label{eq:dh-to-gamma}\delta \tilde h_{ij}(\vec{k})=\sum_{n=1}^{6}  \gamma_n(\vec{k}) \,  {A}^{{(n)}}_{ij}(\hat k)\, ; \hspace{1cm} \delta \tilde \pi^{ij}(\vec{k})=\sum_{n=1}^{6}  \pi_n(\vec{k}) \,  {A}_{{(n)}}^{ij} (\hat k)\, .  \ee
In FLRW spacetimes $\gamma_n$ and $\pi_n$  are called scalar modes for $n=1,2$,  vector modes  for $n=3,4$, and tensor modes for $n=5,6$, due to their properties under rotations around the direction $\hat k$.  We will keep using these names  along this paper. The non-zero Poisson brackets of these  modes are
\bea \label{gammacomm}\{\gamma_n(\v k), \pi_m (\v{k}')\}&=&
\mathcal{V}_0^{-1} \, \delta_{nm} \, \delta_{\vec{k},-\vec{k}'}\, .
\ea
Furthermore, we define
\be\label{eq:df-to-g0}
\gamma_0\equiv \sqrt{4\kappa}\  \delta \t\phi(\vec k),\quad  \pi_0\equiv \sqrt{1/4\kappa}\  \delta \t p_\phi(\vec k),
\ee
and we denote all the degrees of freedom in perturbations as $ \gamma_{\alpha}(\vec k)$ and $  \pi_{\alpha} (\vec k)$ with $\alpha=0,\cdots,6$.

It will be  useful in  the next section to define the products of the shear tensor $\sigma_{ij}$ and  ${A}_{{(n)}}^{ij}$ as
\be\label{eq:sigman}
\sigma_{(n)}(\hat k) \equiv \sigma_{ij} \, {A}_{(n)}^{ij}(\hat k),
\ee
for $n=2,\cdots,6$ ($\sigma_{(1)}$ vanishes, because it is proportional to the trace of  $\sigma_{ij}$). Note however that $\sigma_{(n)}(\hat k) $ is not the Fourier transform of any of the components of  $\sigma_{ij}$.

We can now write the linear constraints $\mathbb S^{(1)}(\v x)$ and $\mathbb V_i^{(1)}(\v x)$ in terms of the variables $\gamma_{\alpha}(\vec k)$ and $\pi_{\alpha}(\vec k)$. For this purpose, we first expand the constraints in Fourier modes $\mathbb S^{(1)}(\v x)=\sum_{\vec k} \tilde {\mathbb S}^{(1)}(\v k)e^{i\, \v{k}\cdot\v{x}}$ and $\mathbb V_i^{(1)}(\v x)=\sum_{\vec k} \tilde  {\mathbb V}_i^{(1)}(\v k)e^{i\, \v{k}\cdot\v{x}}$, and then we replace (\ref{eq:dh-to-gamma}). Explicit expressions are provided in Appendix B of Ref. \cite{aos}.

\subsection{Implementation in {\tt Mathematica}}

We begin by defining the vectors  $\vec k$, $\hat e_1$, and  $\hat e_2$ as follows
\mathIn{DefTensor[kv[-i],\{M3,t\}];}
\mathIn{DefTensor[e1[-i],\{M3,t\},OrthogonalTo->kv[i]];}
\mathIn{DefTensor[e2[-i],\{M3,t\},OrthogonalTo->\{kv[i],e1[i]\}];}
We define the norm of  $\v k$ as
\mathIn{DefTensor[k[],\{M3,t\}];}
\mathIn{AutomaticRules[kv,MakeRule[\{kv[-i]kv[-b]h[i,j],k[]$^{\wedge}$2\}]];}
and we add automatic rules to indicate that $\hat e_1$ and  $\hat e_2$ are unit vectors
\mathIn{AutomaticRules[e1,MakeRule[\{e1[-i]e1[-b]h[i,j],1\}]];}
\mathIn{AutomaticRules[e2,MakeRule[\{e2[-i]e2[-b]h[i,j],1\}]];}
The scalar, vector and tensor modes are defined as follows. First, the symmetric matrix ${A}^{(1)}_{ij}$ is introduced as
\mathIn{DefTensor[A1[-i,-j],\{M3,t\},Symmetric[\{-i,-j\}]];}
and $\gamma_1(\v k)$ and $\pi_1(\v k)$ as
\mathIn{DefTensor[$\gamma$1[],\{M3,t\}];}
\mathIn{DefTensor[$\pi$1[],\{M3,t\}];}
In the same way, we define all other tensors {\tt A2[-i,-j]}, $\ldots$ , {\tt A6[-i,-j]}, and the modes $\gamma${\tt2[]}, $\ldots$ , $\gamma${\tt6[]} and $\pi${\tt2[]}, $\ldots$ , $\pi${\tt6[]}.

We implement the traces and orthogonality properties of these matrices as automatic rules using the command {\tt AutomaticRules}. However, for convenience, we express the matrices  ${A}^{(n)}_{ij}$ in terms of the background metric {\tt h} and the orthogonal vectors {\tt kv[-i]}, {\tt e1[-i]} and {\tt e2[-i]} with the command {\tt MakeRule}. For instance, for the matrices associated to scalar modes, we define 
\mathIn{A1rule=MakeRule[\{A1[-i,-j],h[-i,-j]/Sqrt[3]\}];}
\mathIn{A2rule=MakeRule[\{A2[-i,-j],Sqrt[3/2](kv[-i]kv[-j]/k[]$^{\wedge}$2-h[-i,-j]/3)\}];}
and similarly for the matrices corresponding to vector and tensor modes.
 Finally we define $\gamma_0(\vec k)$ and $\pi_0(\vec k)$:
\mathIn{DefTensor[$\gamma$0[],\{M3,t\}];}
\mathIn{DefTensor[$\pi$0[],\{M3,t\}];}
and their relation with the perturbations of the scalar field and its momentum:
\mathIn{$\gamma$0rule=MakeRule[\{$\delta\phi$[LI[1]],$\gamma$0[]/Sqrt[4$\kappa$]\}];\\
$\pi$0rule=MakeRule[\{$\delta$P$\phi$[LI[1]],$\pi$0[]Sqrt[4$\kappa$]\}];}
We can now introduce the rules that implement  the  decomposition (\ref{eq:dh-to-gamma})
\mathIn{ moderule1=MakeRule[\{$\delta$h[LI[1],-i,-j],$\gamma$1[]A1[-i,-j]+$\gamma$2[]A2[-i,-j]+$\gamma$3[]A3[-i,-j]\\
+$\gamma$4[]A4[-i,-j]+$\gamma$5[]A5[-i,-j]+$\gamma$6[]A6[-i,-j]\}];\\
moderule2=MakeRule[{$\delta$P[LI[1],i,j],$\pi$1[]A1[i,j]+$\pi$2[]A2[i,j]+$\pi$3[]A3[i,j]\\
+$\pi$4[]A4[i,j]+$\pi$5[]A5[i,j]+$\pi$6[]A6[i,j]}];}

Next, we implement the canonical Poisson brackets between $\gamma_{\alpha}(\vec k)$ and $\pi_{\alpha}(\vec k)$ by means of the following function
\mathIn{PoissonBracket[f\_,g\_,q\_List,p\_List]/;Length[q]==Length[p]:=D[f,\{q\}].D[g,\{p\}]\\
-D[f,\{p\}].D[g,\{q\}];\\
PoissonBracket[most\_\_,q:Except[\_List],p\_]:=PoissonBracket[most,\{q\},p];\\
PoissonBracket[most\_\_,p:Except[\_List]]:=PoissonBracket[most,\{p\}];}
If we introduce the following arrays
\mathIn{Q=\{$\gamma$0[],$\gamma$1[],$\gamma$2[],$\gamma$3[],$\gamma$4[],$\gamma$5[],$\gamma$6[]\};\\
PQ=\{$\pi$0[],$\pi$1[],$\pi$2[],$\pi$3[],$\pi$4[],$\pi$5[],$\pi$6[]\};}
one can compute the Poisson brackets of any two phase space functions of perturbations. For instance
\mathIn{PoissonBracket[$\gamma$0[],$\pi$0[],Q,PQ];}
\mathOut{1}
Below we use this function to verify that the Poisson algebra of linear constraints is closed. It is important to keep in mind that the conjugate variable to $ \gamma_{\alpha}(\vec k)$ is  $  \pi_{\alpha} (-\vec k)$. We will take this into account, although we will not implement it explicitly in the notebook for the sake of simplicity. 

We  now define the components $\sigma_{(n)}(\hat k)$ of the shear tensor following Eq. \eqref{eq:sigman} 
\mathIn{DefTensor[$\sigma$2[],\{M3,t\}];}
and similarly  for $\sigma${\tt3[]}, $\ldots$, $\sigma${\tt6[]}. The relation of these quantities and  $\sigma_{ij}$
is implemented via the rule
\mathIn{sheardecomposition=MakeRule[\{$\sigma$[-i,-j],$\sigma$2[]A2[-i,-j]+$\sigma$3[]A3[-i,-j]\\
+$\sigma$4[]A4[-i,-j]+$\sigma$5[]A5[-i,-j]+$\sigma$6[]A6[-i,-j]\}];}
It will be useful in the next sections  to define the following rule
\mathIn{$\sigma$[-i,-j]$\sigma$[i,j]/.sheardecomposition//org;\\
$\sigma$brule=MakeRule[\{$\sigma$b[],\%$^{\wedge}$(1/2)\}];}

We finish this section by writing the linear constraints in Fourier space $ \tilde {\mathbb S}^{(1)}(\v k)$ and $ \tilde  {\mathbb V}_i^{(1)}(\v k)$ and in terms of the  new variables  $\gamma_{\alpha}(\vec k)$ and $\pi_{\alpha}(\vec k)$. For the  diffeomorphism constraints $ \tilde  {\mathbb V}_i^{(1)}(\v k)$ this is implemented by applying the rules 

\mathIn{diffeoa/.MakeRule[\{PD[-b]@$\delta$h[LI[1],-i,-k],I*kv[-j]$\delta$h[LI[1],-i,-k]\}]\\
/.MakeRule[\{PD[-i]@$\delta$P[LI[1],j,k],I*kv[-d]$\delta$P[LI[1],j,k]\}]\\
/.MakeRule[\{PD[-i]@$\delta\phi$[LI[1]],I*kv[-i]$\delta\phi$[LI[1]]\}]/.moderule1\\
/.moderule2/.$\gamma$0rule/.$\pi$0rule//ContractMetric;}
We further simplify the final expression: 
\mathIn{diffeob\,=\,\%/.sheardecomposition/.A1rule/.A2rule/.A3rule/.A4rule\\
/.A5rule/.A6rule//org;}
And now we decompose these constraints in their projections in the directions $\hat k$, $\hat e_1$, and $\hat e_2$
\mathIn{diffeob1=kv[i]diffeob//ToCanonical;\\
diffeob2=e1[i]diffeob//ToCanonical;\\
diffeob3=e2[i]diffeob//ToCanonical;}

We proceed in a similar way with  $ \tilde {\mathbb S}^{(1)}(\v k)$. We do it in several steps. In the first one, we substitute  $\partial_i$ by $i\, k_i$
\mathIn{S1b/.MakeRule[\{PD[-i]@$\delta$h[LI[1],-j,-k],I*kv[-j]$\delta$h[LI[1],-i,-k]\}]\\
/.MakeRule[\{PD[-j]@$\delta$h[LI[1],-i,-k],I*kv[-j]$\delta$h[LI[1],-i,-k]\}]\\
/.MakeRule[\{PD[-i]@kv[i],0\}]//ContractMetric;}
We then apply the SVT decomposition by means of
\mathIn{\%/.moderule1/.moderule2/.$\gamma$0rule/.$\pi$0rule//ToCanonical;}
Finally, we simplify the result 
\mathIn{S1c=\%/.sheardecomposition/.A2rule/.A3rule/.A4rule/.A5rule/.A6rule\\
/.$\sigma$brule//org;}
One can now check that these constraints have vanishing Poisson brackets (modulo the background constraint). For instance, 
\mathIn{PoissonBracket[S1c,diffeob1,Q,PQ]/.bgconstraintrule//org}
\mathOut{0}
where we have evaluated the background constraint on-shell. We have checked that the Poisson brackets with the remaining constraints all vanish, in these cases identically (see the {\tt Mathematica} notebook \cite{ntbk}). Hence, they are  first class constraints, as it must be from the view point of general relativity. 

One can also check that none of the modes $\gamma_{\alpha}$ or their conjugate momenta $\pi_{\alpha}$  commute with these constraints. We show here a couple of examples:
\mathIn{PoissonBracket[$\gamma$0[],S1c,Q,PQ]//org}
\mathOut{$\frac{2\sqrt{\kappa}{\tt P}\phi}{\sqrt{\tilde{\tilde{\tt h}}}}$}
\mathIn{PoissonBracket[$\gamma$1[],S1c,Q,PQ]//org}
\mathOut{$-\frac{\kappa\pi{\tt a}}{\sqrt{3}{\tilde{\tilde{\tt h}}}^{1/3}}$}
Consequently, these variables are not gauge invariant.\footnote{This contrasts with the isotropic case, where tensor modes $\gamma_5$, $\gamma_6$ and their momenta are gauge invariant.}

\section{Gauge invariant variables}\label{sec:gaugeinvarinat}

\subsection{Theory}

We have seven degrees of freedom (per Fourier mode) in configuration variables in the perturbations,  $\gamma_{\alpha}(\vec k)$.  They are  subject to four first class constraints, $\tilde {\mathbb S}^{(1)}(\v k)\approx 0$ and $\tilde  {\mathbb V}_i^{(1)}(\v k)\approx 0$, that  are the generators of gauge transformations for each mode. Each constraint  reduces the number of independent configuration variables by one. In consequence, we are left with $7-4=3$ physical configuration fields, together with their conjugate momenta. We will isolate these degrees of freedom by identifying gauge invariant fields. In the Lagrangian formalism, this was accomplished in Ref.\ \cite{Pereira:2007yy}. In the Hamiltonian framework, the gauge invariant variables are defined  as fields that are left invariant by the gauge flow generated by the linear constraints, or equivalently, that  Poisson-commute with them. 

A conceptually simple and elegant procedure to find gauge invariant variables can be obtained by using the ideas of \cite{gnr}, which were in part applied to FLRW cosmologies in \cite{lang}. The details of this procedure in Bianchi I cosmologies  can be found in  \cite{aos}. In summary, the idea is to find a canonical transformation from $\gamma_{\alpha}(\vec k), \pi_{\alpha}(\vec k)$ to new variables $\Gamma_{\alpha}(\vec k),  \Pi_{\alpha} (\vec k)$ such that the new momenta $\Pi_{\alpha} (\vec k)$ for $\alpha=3,4,5,6$ are proportional to the four constraints $\tilde {\mathbb S}^{(1)}(\v k)$ and $\tilde  {\mathbb V}_i^{(1)}(\v k)$, respectively. More concretely, we demand \be \label{eq:new-Pi}  \Pi_{3} (\vec k)=\frac{1}{|\v k|}\, \tilde{\mathbb{S}}^{(1)}(\vec k)\, , \hspace{0.4cm}  \Pi_{4} (\vec k)=\frac{1}{i \, |\v k|}\, \h k^j\,\tilde{\mathbb{V}}^{(1)}_j(\vec k)\, , \hspace{0.4cm}  \Pi_{5} (\vec k)=\frac{1}{i\, |\v k|}\,  \h e_1^j\,\tilde{\mathbb{V}}^{(1)}_j(\vec k)\, , \hspace{0.4cm}  \Pi_{6} (\vec k)=\frac{1}{i\, |\v k|}\, \h e_2^j \,\tilde{\mathbb{V}}_j^{(1)}(\vec k)\, , \ee where the factors $1/|\v k|$,  with $|\v k|\equiv\sqrt{k_ik^i}\neq 0$, have been introduced for dimensional reasons, and the imaginary unit for convenience. This choice is possible because the constraints are first class, and it can be done globally in the perturbed phase space because of the linearity of the system. If equation (\ref{eq:new-Pi}) is satisfied, then the canonical commutation relations guarantee that $\Gamma_{\alpha}(\vec k)$ and  $\Pi_{\alpha} (\vec k)$ for $\alpha=0,1,2$ Poisson-commute with the constraints, and hence they are gauge invariant. This procedure also guarantees that gauge invariant fields and  pure gauge ones are decoupled in the Hamiltonian (as we will see explicitly below), and hence  dynamics does not mix them. One can then consistently focus attention on gauge invariant perturbations. 
Interestingly, the task of finding such a canonical transformation reduces to solving a Hamilton-Jacobi-like equation for a generating function, and furthermore,  by working in Fourier space, this equation reduces to algebraic equations that are easy to solve in {\tt Mathematica}. More concretely,  we start with a  generating function  in Fourier space 
\be \label{Gform} G(\v k)=\, B^{\alpha \beta}(\v k)\,  \Pi_{\alpha}(\v k) \, \gamma_{\beta}(\v k)+A^{\alpha \beta}(\v k)\, \gamma_{\alpha}(\v k) \gamma_{\beta}(\v k) \, ,
\ee
 that we choose to be of type 2---i.e.\  it depends on old variables $ \gamma_{\alpha}$ and new momenta $ \Pi_{\alpha}$---and from which the rest of variables are given by
\begin{equation}\label{eq:pi-Gamma-G}
  \pi_{\alpha} (\vec k)= \f{\partial G( \gamma_{\beta},\,  \Pi_{\beta})}{\partial \gamma_{\alpha}(\vec k)} ,
 \hspace{2cm}
  \Gamma_{\alpha} (\vec k)= \f{\partial G( \gamma_{\beta},\,  \Pi_{\beta})}{\partial \Pi_{\alpha}(\vec k)} .
\end{equation}
where $B^{\alpha\beta}$ and $A^{\alpha \beta}$ are matrices whose components   depend on background variables, but not on perturbations, and furthermore  $A^{\alpha \beta}$ is symmetric. Equations (\ref{eq:pi-Gamma-G}) provide then a set of algebraic relations for the  $77$ unknown coefficients $B^{\alpha\beta}$ and $A^{\alpha \beta}$, although only   38 of them are  independent. Hence, there is freedom in choosing gauge invariant variables.  As mentioned above, we want to choose  gauge invariant fields that in the isotropic limit reduce to the familiar comoving curvature perturbations and the two tensor modes. Indeed,  gauge invariant fields $\Gamma_{\alpha}(\vec k)$ satisfying this property can be identified by inspection, just by looking at the Poisson brackets of $\gamma_{\alpha}$ and the linear constraints. They are
\begin{eqnarray}
  \Gamma_0(\vec k)\, &=&   \gamma_0\, +\, 
\f{\sqrt{\kappa}\,p_{\phi}}{\sqrt{1/6}\, \kappa\,a \, \pi_a\, 
 +\, a^3\, \sigma_{(2)} }\l(\sqrt{2} \, \gamma_1\, -\,  \gamma_2 \r)\, ,
\label{eq:G0}
\\
  \Gamma_1(\vec k)\, &=& \gamma_5\, +
\ \f{\,a^2\,\sigma_{(5)}}{\sqrt{1/6}\, \kappa\,\pi_a\, 
 +\,a^2\, \sigma_{(2)} }\ \l(\sqrt{2}\,  \gamma_1\, -\,  \gamma_2 \r)\, ,
\label{eq:G1}
\\
   \Gamma_2(\vec k)\, &=& \gamma_6\, +
\ \f{\,a^2\,\sigma_{(6)}}{\sqrt{1/6}\, \kappa\,\pi_a\, 
 +\, \,a^2\, \sigma_{(2)} }\ \l(\sqrt{2}\,  \gamma_1\, -\,  \gamma_2 \r)\, ,
\label{eq:G2}
\end{eqnarray}
This will be our choice of gauge invariant fields. Other choices are possible, and all of them can be derived by using the companion notebook \cite{ntbk}. In the isotropic limit $\sigma_{(n)}\to 0$,  $ \Gamma_1$ and $ \Gamma_2$  reduce to the familiar two polarizations of transverse and traceless tensor modes,  and
$\Gamma_0$  becomes proportional to the comoving curvature perturbation  ${\mathcal{R}}(\vec k)\equiv\frac{1}{\sqrt{4\kappa}}\, \frac{a}{z}\Gamma_0$, where $z=-\frac{6}{\kappa}\frac{p_{\phi}}{\pi_a}=\frac{\dot \phi}{H}\, a$.

Identifying the form of the new gauge invariant variables $\Gamma_{\alpha}$ in terms of $\gamma_{\alpha}$ will simplify the computation of $G(\v k)$ (for instance, this determines the value of some of the coefficients $B^{\alpha\beta}$), but it is important to emphasize that the method, and therefore its implementation in the algorithm reported in our manuscript, allows us to work with gauge invariant variables and pure gauge ones in a systematic way, without having to identify them a priori. To our knowledge, numerical tools meeting all these requirements (i.e. handle efficiently complicated phase space functions and keep the construction as general as possible) are not publicly available, at least in the context of cosmological perturbation theory or similar settings.

In addition, we will demand (see next section)  the Hamiltonian to be  ``diagonal'' in the new configuration variables and momenta, i.e.\ we will eliminate ``cross terms'' of the type  $\Gamma_{\alpha}\Pi_{\beta}$. This aesthetic condition will impose further restriction in the coefficients $A^{\alpha \beta}$'s.  The rest of free coefficients  can be equated to zero for simplicity. Once all the coefficients $A^{\alpha \beta}$ and $B^{\alpha \beta}$ are specified, the form of the conjugate momenta $\Pi_{\alpha}$ for $\alpha=0,1,2$ are obtained from (\ref{eq:pi-Gamma-G}).

\subsection{Implementation in {\tt Mathematica}}

We  summarize here the main steps of the code \cite{ntbk}. We start defining the new variables $\Gamma_{\alpha}(\vec k), \Pi_{\alpha}(\vec k)$:
\mathIn{DefTensor[$\Gamma$0[],\{M3,t\}];\\
DefTensor[$\Pi$0[],\{M3,t\}];}
and similarly for  $\alpha=1,\,\ldots\,,6$. 

As explained above, rather than solving all the equations that constrain the coefficients $A^{\alpha\beta}$ and $B^{\alpha\beta}$, 
we simplify the calculation by identifying suitable gauge invariant  variables (see equation (\ref{eq:G2}) above). 
These variables are named in the code as {\tt $\Gamma$0new}, {\tt $\Gamma$1new} , {\tt $\Gamma$2new}: 
\mathIn{$\Gamma$0new=$\gamma$0[]+(6\,Sqrt[2\,$\kappa$]\,P$\phi$\,$\gamma$1[])/((Deth[]$^{\wedge}$(1/6))\,(Sqrt[6]\,$\kappa$\,$\pi$a[]\\+6\,(Deth[]$^{\wedge}$(1/6))$^{\wedge}$2\,$\sigma$2[]))-(6\,Sqrt[$\kappa$]\,P$\phi$[]\,$\gamma$2[])/((Deth[]$^{\wedge}$(1/6))\,(Sqrt[6]\,$\kappa$\,$\pi$a[]\\
+6\,(Deth[]$^{\wedge}$(1/6))$^{\wedge}$2\,$\sigma$2[]));}
and similarly for {\tt $\Gamma$1new} , {\tt $\Gamma$2new}. With this choice we can determine some of the coefficients $B^{\alpha\beta}$. The remaining coefficients in $B^{\alpha\beta}$, namely those with $\alpha,\beta = 3,\,4,\,5,\,6$ are written as unknowns {\tt CI[]}, {\tt EI[]}, {\tt FI[]} and {\tt JI[]}, with {\tt I}=$\{0,\ldots,6\}$. For example, we define
\mathIn{DefTensor[C0[],{M3,t}];\\
DefTensor[E0[],{M3,t}];\\
DefTensor[F0[],{M3,t}];\\
DefTensor[J0[],{M3,t}];}
and similar definitions for the remaining unknowns. We use these coefficients to define
\mathIn{$\Gamma$3new=$\gamma$0[]C0[]+\ldots+$\gamma$6[]C6[];\\
$\Gamma$4new=$\gamma$0[]E0[]+\ldots+$\gamma$6[]E6[];\\
$\Gamma$5new=$\gamma$0[]F0[]+\ldots+$\gamma$6[]F6[];\\
$\Gamma$6new=$\gamma$0[]J0[]+\ldots+$\gamma$6[]J6[];
}
The coefficients $A^{\alpha\beta}$ are denoted by {\tt AIJ[]}, with {\tt I,J}=$\{0,\ldots,6\}$. For instance, we define
\mathIn{
DefTensor[A00[],{M3,t}];}
and similarly for the remaining ones.

Using these coefficients and the definitions of new variables in terms of them, we introduce the definition of the generating function as
\mathIn{G=$\Gamma$0new\,$\Pi$0[]+$\Gamma$1new\,$\Pi$1[]+$\ldots$+$\Gamma$5new\,$\Pi$5[]+$\Gamma$6new\,$\Pi$6[]+A00[]\,$\gamma$0[]\,$\gamma$0[]\\
+2\,A01[]\,$\gamma$0[]\,$\gamma$1[]+$\ldots$+2\,A56[]\,$\gamma$5[]\,$\gamma$6[]+A66[]\,$\gamma$6[]\,$\gamma$6[]//NoScalar;}

We now follow the strategy described above. Namely, we first obtain an expression for the old momenta $\pi_{\alpha}$ in terms of $\gamma_{\alpha}$ and $\Pi_{\beta}$ by taking  derivative  of the generating function with respect to  $\gamma_{\alpha}$. These expressions can then be substituted in the scalar and vector constraints, to express them in terms of old configuration variables $\gamma_{\alpha}$ and new momenta $\Pi_{\beta}$.  By equating  these constraints to  $\Pi_{\alpha}$ for $\alpha=3,4,5,6$ as indicated in (\ref{eq:new-Pi}), we obtain 44 algebraic equations for the coefficients $A^{\alpha\beta}$, out of which 38 are independent. Some of the remaining coefficients can be solved by demanding that the new Hamiltonian does not contain  cross terms between new momenta and new configuration variables. Finally, the remaining coefficients are set to zero for the sake of simplicity. See the notebook \cite{ntbk} for details.

\section{Dynamics of gauge invariant perturbations}\label{sec:2nd-H}

\subsection{Theory}

The dynamics of perturbations is generated by the second order scalar constraint $\int \d^3x\,  N \, \mathbb{S}^{(2)}(\vec x)$. Let us notice that the second order vector constraints $\mathbb{V}_{i}^{(2)}$ do not contribute since the homogeneous part of the shift $N_i$ is zero and the next contribution would come from $\delta N_i(\vec x)\, \mathbb{V}_{i}^{(2)}$, which is third order in perturbations. If we expand the fields inside $\mathbb{S}^{(2)}(\vec x)$ in Fourier modes, the expression for $\int \d^3x\,  N \, \mathbb{S}^{(2)}(\vec x)$ can be written as $\sum_{\vec {k}} N \, \tilde{\mathbb{S}}^{(2)}(\vec k)$, where $\tilde{\mathbb{S}}^{(2)}(\vec k)$ is quadratic in perturbations. In each term, one perturbation is evaluated at $\vec k$ and the other at $-\vec k$.

From the expression for $\tilde{\mathbb{S}}^{(2)}(\vec k)$ in terms of $\delta \tilde{h}_{ij}$ and $\delta \tilde{\pi}^{ij}$, we obtain a Hamiltonian for the new variables $\Gamma_\alpha(\vec k)$ and $\Pi_{\alpha}(\vec k)$, by first implementing the change from $\delta \tilde{h}_{ij}(\vec k),\,  \delta \tilde{\pi}^{ij}(\vec k)$ to $\gamma(\vec k),\, \pi(\vec k)$ and then from the later to $\Gamma_\alpha(\vec k), \Pi_{\alpha}(\vec k)$. One has to keep in mind that these transformations involve coefficients that depend on functions of the Bianchi I background geometry, and therefore they have to be understood as time-dependent quantities. This means that the final Hamiltonian is equal to the original one in new variables, plus the time derivative of the generating function of the canonical transformation, where the time derivative only affects the background functions. The first  canonical transformation  can be implemented by the following generating function 
\begin{equation} \label{gform}
G_{\gamma}(\v k) = - \delta \tilde{\pi}^{ij}(\v k) \sum_{n=1}^6A^{(n)}_{ij}(\v k) \gamma_n(\v k)\, ,  
\end{equation}
where we have chosen it to be of third type (i.e.\, it depends on new configuration variables and old momenta). Recall that the matrices  $A^{(n)}_{ij}(\v k)$ depend on time. The second canonical transformation from $\gamma_\alpha(\vec k),\, \pi_\alpha(\vec k)$ to $\Gamma_\alpha(\vec k), \Pi_{\alpha}(\vec k)$ is defined by the generating function (\ref{Gform}).  

After implementing these transformations, one can check that  gauge invariant fields decouple from pure gauge ones, and one obtains the following Hamiltonian for the former (see \cite{aos} for further details)
\be \label{Ham2}\mathcal{H_{\rm pert}}=\frac{N(t)\,{\cal V}_0}{2\, a(t)}\, \sum_{\vec k}\sum_{\mu,\mu'=0}^2\, \left[ \frac{4\kappa}{a^2(t)}\, \delta_{\mu,\mu'}\, |\Pi_{\mu}(\vec k)|^2+\, \frac{a^2(t)}{4\kappa}\, \Big( \delta_{\mu,\mu'}\, k^2(t)+\, {\cal U}_{\mu \mu'}(t,\vec k)\Big)\, \Gamma_{\mu}(\vec k)\bar \Gamma_{\mu '}(\vec k)\right]\, ,\ee
where $\delta_{\mu,\mu'}$ is the Kronecker delta, and $k^2(t)\equiv a^2(t) \, k^ik_j=a^2(t)\, \left(\frac{k_1^2}{a_1^2(t)}+\frac{k_2^2}{a_2^2(t)}+\frac{k_3^2}{a_3^2(t)}\right)$. If we choose $N=1$, this Hamiltonian  generates evolution in proper time $t$, and in conformal time if $N=a$.  The (time-dependent) effective potentials $ {\cal U}_{\mu\mu'}$ are symmetric in $\mu$ and $\mu'$ and they become diagonal (i.e.\,  proportional to $
\delta_{\mu\mu'}$)  in the isotropic limit. But in presence of  anisotropies, they couple  gauge invariant perturbations among themselves. They are explicitly  written in Appendix \ref{app}.

The equations of motion are (we use cosmic time) %
\bea \label{perteqH} \dot \Gamma_{\mu}(\vec k)&=&\{\Gamma_{\mu}(\vec k),\mathcal{H_{\rm pert}}\}=\frac{4\kappa}{a^3}\, \Pi_{\mu}(\vec k)\, ,\nonumber \\ 
 \dot \Pi_{\mu}(\vec k)&=&\{\Pi_{\mu}(\vec k),\mathcal{H_{\rm pert}}\}=-\frac{a}{4\kappa}\,\sum_{\mu'=0}^2\,  (\delta_{\mu\mu'}\, k^2+ {\cal U}_{\mu \mu'})\, \Gamma_{\mu'}(\vec k)\, . \ea
Combining these equations into second order differential equations we obtain
\be \label{eqginper}\ddot \Gamma_{\mu}+3\, H\, \dot \Gamma_{\mu}+\frac{k^2}{a^2}\,  \Gamma_{\mu}+\frac{1}{a^2}\, \sum_{\mu'=0}^2\, {\cal U}_{\mu\mu'}\, \Gamma_{\mu'}=0\, . \ee
This is a  set of three coupled, second order, ordinary differential equations for each wavevector $\vec k$, and they reduce to the familiar (decoupled) equations for scalar and tensor perturbations in the isotropic FLRW limit

\subsection{Implementation in {\tt Mathematica}}

We start with the expression $\sum_{\vec {k}} N \, \tilde{\mathbb{S}}^{(2)}(\vec k)$ in terms of  $\delta \tilde{h}_{ij}(\vec k)$ and $\delta \tilde{\pi}^{ij}(\vec k)$. As mentioned above, each term $\tilde{\mathbb{S}}^{(2)}(\vec k)$  is quadratic in perturbations, with one field evaluated at $\vec k$ and the other at $-\vec k$. In our {\tt Mathematica} notebook this will remain implicit, since the code will be significantly simpler in this way. 

We first obtain an expression for  $\tilde{\mathbb{S}}^{(2)}(\vec k)$ in terms of  $\delta \tilde{h}_{ij}(\vec k)$ and $\delta \tilde{\pi}^{ij}(\vec k)$:
\mathIn{S2a=SeriesCoefficient[S,2];}
\mathIn{S2a/.MakeRule[{PD[-i]@PD[-j]@$\delta$h[LI[1],-k,-l],-kv[-i]kv[-j]$\delta$h[LI[1],-k,-l]}]\\
/.MakeRule[{PD[-i]@$\delta\phi$[LI[1]]PD[-j]@$\delta\phi$[LI[1]],kv[-i]kv[-j]$\delta\phi$[LI[1]]$\delta\phi$[LI[1]]}]\\
/.MakeRule[{PD[-i]@$\delta$h[LI[1],-j,-k] PD[-d]@$\delta$h[LI[1],-l,-m],kv[-i]kv[-l]\\
$\delta$h[LI[1],-j,-k]$\delta$h[LI[1],-l,-m]}]/.MakeRule[{h[i,j]kv[-i]kv[-j],k[]$^\wedge$2}]\\
/.MakeRule[{PD[-j]@PD[-l]@$\delta$h[LI[2],-i,-k],0}]\\
/.MakeRule[{PD[-k]@PD[-l]@$\delta$h[LI[2],-i,-j],0}]/.bgmomrule//org;}
where, as before, we have replaced spatial derivatives by $i\vec k$. The second step is to move from $\delta \t{\phi}(\vec{k}),\, \delta \t{p}_{\phi}(\vec{k}),\, \delta \tilde{h}_{ij}(\vec k),\,  \delta \tilde{\pi}^{ij}(\vec k)$ to $\gamma_\alpha(\vec k),\, \pi_\beta(\vec k)$ :
\mathIn{\%/.moderule1/.moderule2/.$\gamma$0rule/.$\pi$0rule//org;}
and we also decompose the shear in its components $\sigma_{(n)}$:
\mathIn{\%/.sheardecomposition//org;}
We further simply our final result with
\mathIn{S2b=\%/.A2rule/.A3rule/.A4rule/.A5rule/.A6rule/.$\sigma$brule//org;}
Finally, we  want to write the Hamiltonian in terms of the new  momenta $\Pi_{\alpha}$ and configuration variables $\Gamma_{\alpha}$. In order to do so, we introduce in the notebook the expressions for old configuration variables as functions of new ones as {\tt $\gamma$0old1}, $\ldots$ , {\tt $\gamma$6old1}. Similarly, we introduce expressions for old momenta in terms of new momenta and old configuration variables which are denoted by {\tt $\pi$0old1}, $\ldots$ , {\tt $\pi$6old1}. One can easily combine these two sets of expressions to replace old configuration variables and momenta in any expressions by new ones. Thus, we obtain
\mathIn{S2b/.MakeRule[{$\pi$0[],$\pi$0old1}]/.MakeRule[{$\pi$1[],$\pi$1old1}]/.MakeRule[{$\pi$2[],$\pi$2old1}]\\
/.MakeRule[{$\pi$3[],$\pi$3old1}]/.MakeRule[{$\pi$4[],$\pi$4old1}]/.MakeRule[{$\pi$5[],$\pi$5old1}]\\
/.MakeRule[{$\pi$6[],$\pi$6old1}];\\
S2c=\%/.MakeRule[{$\gamma$0[],$\gamma$0old1}]/.MakeRule[{$\gamma$1[],$\gamma$1old1}]/.MakeRule[{$\gamma$2[],$\gamma$2old1}]\\
/.MakeRule[{$\gamma$3[],$\gamma$3old1}]/.MakeRule[{$\gamma$4[],$\gamma$4old1}]/.MakeRule[{$\gamma$5[],$\gamma$5old1}]\\/.MakeRule[{$\gamma$6[],$\gamma$6old1}];}
As explained above, the final Hamiltonian is obtained from this expression by adding the time derivative of the generating functions (\ref{gform}) and (\ref{Gform}). We denote by {\tt dG$\gamma$dt1} and {\tt dGdt2} their time derivatives in the companion notebook, respectively. The addition of these terms requires some additional simplifications. Concretely, we apply the following rule in several intermediate steps
\mathIn{bgconstraintrule=MakeRule[{V[$\phi$[]],-1/(2$\kappa$)($\sigma$2[]$^{\wedge}$2+$\sigma$3[]$^{\wedge}$2+$\sigma$4[]$^{\wedge}$2+$\sigma$5[]$^{\wedge}$2+$\sigma$6[]$^{\wedge}$2)\\
-1/Deth[]$^{\wedge}$(1/2)(P$\phi$[]$^{\wedge}$2/(2Sqrt[Deth[]])-($\kappa$$\pi$a[]$^{\wedge}$2)/(12Deth[]$^{\wedge}$(1/6)))}];}
It uses  the background constraint to  replace the potential $V(\phi)$ of the scalar field by the other background variables. We get
\mathIn{S2d = S2c+dGdt1+dG$\gamma$dt2/.bgconstraintrule;}
This expression can be further simplified. On the one hand, it contains terms proportional to the new momenta $\Pi_3$, $\ldots$ , $\Pi_6$, which, by construction, are constrained to vanish. Therefore, we set them equal to zero with the rule
\mathIn{S2e=S2d/.MakeRule[{$\Pi$3[],0}]/.MakeRule[{$\Pi$4[],0}]/.MakeRule[{$\Pi$5[],0}]/.MakeRule[{$\Pi$6[],0}]\\
//org;}

We have verified that there is no coupling between gauge invariant and pure gauge variables, once the background constraint is also imposed. Therefore, we can focus on the part of the Hamiltonian that contains only gauge-invariant fields. Nevertheless, this part still contains cross terms between $\Pi_\mu$ and $\Gamma_\mu$, with $\mu=0,1,2$. Fortunately, and as discussed above, the generating function $G(\vec k)$ still has some free parameters that can  be fixed by requiring that these terms vanish. Concretely, we require the coefficients multiplying the cross terms {\tt ($\Gamma$0[] $\Pi$0[])}, {\tt ($\Gamma$1[] $\Pi$0[])}, {\tt ($\Gamma$2[] $\Pi$0[])}, {\tt ($\Gamma$1[] $\Pi$1[])}, {\tt ($\Gamma$1[] $\Pi$2[])} and {\tt ($\Gamma$2[] $\Pi$2[])}, to vanish. These conditions fix the coefficients {\tt A01[], A05[], A06[], A56[], A12[], A26[]} in (\ref{Gform}). As a bonus, this automatically guarantees that the remaining cross terms {\tt ($\Gamma$0[] $\Pi$1[])}, {\tt ($\Gamma$1[] $\Pi$0[])}, {\tt ($\Gamma$0[] $\Pi$2[])}, {\tt ($\Gamma$2[] $\Pi$0[])}, {\tt ($\Gamma$1[] $\Pi$2[])} and {\tt ($\Gamma$2[] $\Pi$1[])} all vanish. As mentioned before, we set the remaining free parameters to zero, in order to have a  Hamiltonian as simple as possible. 

Finally, in order to focus on gauge invariant fields, we set pure gauge variables to zero in the code. This yields an expression for the Hamiltonian that we denote by {\tt H2f}, which is further simplified by means of the rule
\mathIn{H2g=H2f/.bgconstraintrule//org;}
This final Hamiltonian can  be simplified and written in the form  given in (\ref{Ham2}). Equations of motion for the gauge invariant variables can be derived straightforwardly from it.

\section{Discussion}\label{sec:concl}

We have described in this paper the main steps of a computer code written in the symbolic language of {\tt Mathematica}, and made publicly available in \cite{ntbk}, that derives gauge invariant linear perturbations in Bianchi type I cosmological spacetimes in a Hamiltonian or phase space approach. In this formulation, gauge invariant linear perturbations are defined by  fields that Poisson-commute with the linear constraints of the theory, and they can be found systematically by solving a canonical transformation that identifies some of the new momenta with the constraints. We have described in detail the implementation of this procedure in {\tt Mathematica}. Our code provides an efficient tool to explore different choices of linear gauge invariant fields, and to derive the equations of motion they satisfy. It can also be used to work with gauge dependent variables, after choosing a gauge, and to relate physical observables written in different gauges.
Furthermore, we have complemented this analysis with a computer code, based on the C programing language, available in  \cite{num-lib}, that {\em solves} the equations of motions and computes observables that can be compared with current and future data from the cosmic microwave background (CMB). Our computer codes should be of great utility for researchers interested in cosmological perturbations in Bianchi spacetimes and their consequences for the CMB, as well as in isotropic FLRW, which is obtained by simply putting the anisotropies to zero. Our theoretical and numerical analysis can also  be of interest for pedagogical purposes, since it provides a step by step, guided way of implementing  the theory of gauge invariant cosmological perturbations in a computer. 



\acknowledgments{We have benefited from  discussions with Abhay Ashtekar,  Mar Bastero-Gil, Brajesh Gupt, Guillermo A. Mena Marug\'an, Jorge Pullin, Parampreet Singh and Edward Wilson-Ewing. 
This work is supported by the NSF CAREER grant PHY-1552603, and from the Hearne Institute for Theoretical Physics.  We acknowledge the use of high performance computing resources provided by Louisiana State University (http://www.hpc.lsu.edu), Baton Rouge, U.S.A. }


\appendix
\section{Hamiltonian of gauge invariant perturbations}\label{app}
The potentials $\mathcal{U}_{\mu\mu'}$ in the Hamiltonian (\ref{Ham2}) that generates the dynamics of gauge invariant perturbations are given by the following expressions
{\bea
\, {\cal U}_{00}&=&a^2\, V_{\phi\phi}\, - \frac{2 \kappa \, \pp^2 {\cal F}_2}{a^3}  + 2 \kappa  \, {\cal F}_1  \left(-\frac{\kappa\,\pp^2\,p_a}{3a^{5}}\, + \,2 \,V_{\phi}\, \pp\right),\\ \nonumber
{\cal U}_{01}&=&{\cal U}_{10} \,= \,\frac{2\sqrt{\kappa}}{a^2}\left(-a^2\,\pp\, \sigma_{(5)} \, {\cal F}_2  + a^5V_{\phi} \, \sigma_{(5)}\, {\cal F}_1 - a^2\,\pp \, {\cal G}_{5}\,  {\cal F}_1\, +\, \frac{\kappa}{6}\,\pp\,p_a\,\sigma_{(5)}\,{\cal F}_1\right)\,,\\ \nonumber
{\cal U}_{02}&=&{\cal U}_{20}\,= \,\frac{2\sqrt{\kappa}}{a^2}\left(-a^2\,\pp\, \sigma_{(6)} \, {\cal F}_2  + a^5\,V_{\phi} \, \sigma_{(6)}\, {\cal F}_1 - a^2\,\pp \, {\cal G}_{6}\,  {\cal F}_1\, +\, \frac{\kappa}{6}\,\pp\,p_a\,\sigma_{(6)}\,{\cal F}_1\right)\,,\\ \nonumber
{\cal U}_{12}&=&{\cal U}_{21}\,=\, 2\,\sigma_{(5)}\,\sigma_{(6)}\,\left( a^2 -\, a^3\,{\cal F}_2\,+\, \frac{2}{3}\,\kappa\,a\,p_a\,{\cal F}_1\right) - \left( \, 2\,a^3\,\sigma_{(6)}\,{\cal G}_{5}\, +\, 2\,a^3\,\sigma_{(5)}\,{\cal G}_{6}\right)\,{\cal F}_1\, \\ \nonumber
{\cal U}_{22}\, &=&\,- 2 \,a^2\, \sigma_{(5)}^2\, +\,\frac{\kappa p_a \,\sigma_{(2)}}{\sqrt{6}}\,-\,a^2\,\sqrt{\f{2}{3}}{\cal G}_2\, +\,\frac{4}{3}\,\kappa\,a\,p_a\,\sigma_{(6)}^2\,{\cal F}_1\,-\,4 \,a^3 \,\sigma_{(6)}\, {\cal F}_1\, {\cal G}_6\,-\,2\,a^3\, \sigma_{(6)}^2\, {\cal F}_2\,, \ea}
 with $V_{\phi}\equiv dV/d\phi$, $V_{\phi\phi}\equiv d^2V/d\phi^2$, and
\bea
{\cal F}_1\, &=&\, \frac{-\frac{\kappa p_a}{2a^3}\, +\,\sqrt{\frac{3}{2}} \,\frac{\sigma_{(2)}}{a}}{
      2\kappa\rho\,+\,\sigma_{(3)}^2\,+\, \sigma_{(4)}^2+\, \sigma_{(5)}^2\,+\, \sigma_{(6)}^2},\\ \nonumber
{\cal F}_2\, &=&\frac{\frac{3\kappa \, V}{a}\, - \,\frac{\kappa^2 p_a^2}{3a^5}\,+\,\frac{\kappa p_a\sigma_{(2)}}{2\sqrt{6}a^3}\, +\, \sqrt{\frac{3}{2}}\frac{{\cal G}_{2}}{a}\,-\,{\cal F}_1\left[\frac{\kappa^2 \pp^2p_a}{a^8}\,+\,2\,\sigma_{(3)}\, {\cal G}_3\,+\,2\,\sigma_{(4)}\, {\cal G}_4+\, 2\,\sigma_{(5)}\, {\cal G}_5\,+\, 2\,\sigma_{(6)}\, {\cal G}_6)\right]}{
    2\kappa\rho\,+\,\sigma_{(3)}^2\,+\, \sigma_{(4)}^2+\, \sigma_{(5)}^2\,+\, \sigma_{(6)}^2},\\ \nonumber
{\cal G}_2 &=& \frac{\kappa p_a\sigma_{(2)}}{2\,a^2}\, -\, \sqrt{\frac{3}{2}}\left(\sigma_{(3)}^2 \,+\, \sigma_{(4)}^2\right),\\ \nonumber
{\cal G}_3 &=& \frac{\kappa\, p_a\, \sigma_{(3)}}{2\,a^2}\,  +\, \frac{1}{\sqrt{2}} \left(\sqrt{3}\sigma_{(2)} \sigma_{(3)} - \sigma_{(3)} \sigma_{(5)} - \sigma_{(4)} \sigma_{(6)}\right),\\ \nonumber
{\cal G}_4 &=&  \frac{\kappa p_a\sigma_{(4)}}{2\,a^2} + \frac{1}{\sqrt{2}} \left(\sqrt{3}\sigma_{(2)} \sigma_{(4)} + \sigma_{(4)} \sigma_{(5)} - \sigma_{(3)} \sigma_{(6)}\right)\,, \\ \nonumber
{\cal G}_5 &=& \frac{\kappa p_a\sigma_{(5)}}{2\,a^2}\, +\, \frac{1}{\sqrt{2}}(\sigma_{(3)}^2 - \sigma_{(4)}^2),\\ \nonumber
{\cal G}_6 &=& \frac{\kappa p_a\sigma_{(6)}}{2\,a^2} + \sqrt{2}\,\sigma_{(3)}\sigma_{(4)}.
\ea
Note that the expressions above have an implicit  dependence in $\vec k$ coming from $\sigma_{(n)}(\vec k)$. 




\end{document}